*A Case Study on Data Acquisition Systems:*
*Relevance to Renewable Energy Technologies*

Chito A. Petilla
University of San Carlos – Talamban Campus
Eastern Visayas State University – Main Campus

**Abstract**

Multiple advantages had been identified with the integration of data acquisition into any existing system configuration and implementation. Using data acquisition as a support into a monitoring system has not only improved its overall performance and reliability but also lowered its operational and maintenance cost because of its real-time data collection from node sensors.

As renewable energy needs to be sustainable for it to fully support the energy demand of communities, its management and control still needs to be improved and enhanced. Smart systems are considered the next generation technological improvement of any system that exists. It is the prelude to autonomous systems from industrial applications to home automation. Data acquisition is only a part of these smart systems that help in the remote management and control of these devices. Remote monitoring functionality enhances the operation and reliability which help in making proactive decisions during critical situations and circumstances.

Even with data acquisition enhancements, there is still room for improving its implementation regarding data security and privacy and accuracy of information being exchanged between nodes. Current technological advancements have already shown promising results and have widen its utilization spectrum by covering almost any field of specialization. With increasing implementation and design complexity that comes with its enhancements, challenges and issues are also faced that needs to be addressed and considered to mitigate the effects of such.

## 1  Introduction

Integration of data acquisition into an existing system enables remote management and control of inaccessible installations. It facilitates the real-time monitoring, fault and anomaly detection, error control, automation and perform predictive maintenance to the monitored system. By 2015, network connected devices energy market exceeded $6.8B and is projected to be more than $56.5B by 2023, with a compounded annual growth of 15.5% between 2016 up to 2023 [1]. In 2024, there will be about 20 billion connected objects around the world [2]. Through this connectivity in data transmission networks, cloud infrastructures such as cloud computing, edge computing and quantum computing are used to analyze and visualize real-time information for efficient data processing and analysis.

With the integration of data acquisition into deployment of remote monitoring systems, it has introduced opportunities in lowering operational and maintenance cost at the same time increasing reliability and efficiency of the systems [3]. Even with its rapid development and adoption, this technology is still in its refinement stage because of the identified security challenges that it presents. These challenges can be technical and technological challenges, lack of budget and knowledge, information security and data protection challenges, scalability, standardization issues, legal and regulatory security and interoperability of the network environment [4]. Networked devices have shown to be promising in terms of resource management efficiency. These current trends in technological advancement have also identified emergence of new technological attacks that has been prevalent in recent years [5].

Among the various sectors that had taken advantage of becoming intelligent systems are healthcare, utilities, manufacturing, transportation and housing [6]. With the technological advancement of interconnecting the different devices globally, it has made a digital revolution in the real world. It allowed medical equipment, wearable technology, surveillance cameras, digital devices, home appliances, environmental and weather sensors to be used in improving the lives of the people and the environment.

Different techniques of data gathering from remote node devices and sensors are discussed in this study. Included also in this study are identifying the current challenges and concerns being faced in its implementation likewise with the opportunities that comes with integrating data acquisition into a system.

## 2 Data Acquisition Systems: Its Technologies and Methods

### 2.1 Sensor Technologies

Data acquisition systems through the use of sensors plays a very important role in effectively managing and controlling remote systems for the improved performance and increased reliability of these monitored systems. Information received from these remote systems are used to evaluate the performance of the system for optimization and maintenance purposes. Responding to a fault as quickly as possible is critical in any large-scale systems and the ability to react to potential system failure is directly related to rapid detection of a faulty system [7]. Slow response to these system failures can cause significant losses in both efficiency and economically.

Data collection and analysis have been optimized with the digitization of traditional data acquisition systems, through the elimination of traditional single node data acquisition failures such as sensor faults and data anomalies [8]. With this development, processing for evaluating and analyzing massive influx of real-time data sets from sensor nodes have improved the systems monitoring and control capabilities [9]. This process has also enhanced the systems' predictive capabilities, enable proactive maintenance and decision-making to prevent potential failures and optimize resource allocation.

### 2.2 Solar Photovoltaic Data Acquisition Systems

Solar photovoltaic energy monitoring systems have been made to collect and analyze power generation data, and integrating node devices with it allows for performance prediction and reliable power output maximizing solar energy usage [10]. Global installations of PV systems have significantly grown over the years, having commissioned an almost 175GW worldwide in 2021 and with cumulative installed capacity that reached 942GW at the end of 2021 [11]. Energy management technology is used to maximize economical utilization without sacrificing overall energy efficiency.

With the need of photovoltaic systems of sophisticated real-time monitoring capabilities for proper maintenance, operation and control, for it to minimize and avoid unwanted electric power disruptions, swift failure detection and reduce operating and maintenance cost [12]. Undetected solar PV system failures have a very negative economic effect thus making monitoring systems crucial to the daily operation and performance of renewable energy generation systems.

Even without the availability of network for data gathering, it is still equally important to retrieve operational parameters and conditions of a system while in operation. A local datalogger can be used to measures electric and climactic and meteorological characteristic of a remote smart energy system installed on sites that have no access to wired data network or electrical grid [13]. Data from sensor readings can be stored in local storage and can be retrieve for analysis and evaluation.

Solar photovoltaic system needs an efficient battery charge controller that balances photovoltaic energy flow to the battery storage and to the load thus ensuring the optimal utilization of generated PV energy and appropriate battery charging [14]. This requirement needs the system to have an energy management algorithm to ensure battery safety by considering weather conditions and load demand variations through its system's mode of operation. Collecting appropriate and accurate information from sensors are crucial in the data analysis of the algorithm to execute.

Flexible solar modules innovation has also maximized and optimize energy resource management thru its ability to adapt different surface conditions. These modules are made from thin-film silicon materials which are perfect for non-flat surfaces and have the flexibility advantage to be installed on any surface. Understanding the capability of energy capture of this device is crucial to determine the viability of using these modules compared to using the conventional solar modules [15]. Using flexible solar modules even on non-flat surface can generate energy comparable to a traditional rigid module based on data collected from node sensors.

In a solar photovoltaic system, the life of the batteries primarily depends also on the solar charge controller settings which manages the energy balance of the whole assembly [16]. Sensors are used to measure and analyze the static and dynamic characteristics of the batteries that allowed to dynamically adjust the available maximum battery capacity during operation. In this process, the capability of the energy storage is maximized and the operation of the whole system is optimized.

## 3 Smart Systems: Applications

Data acquisition has been integrated into different applications and installations. It has enhanced the living standards of everyone, where anybody and anything can be inter-connected to the network and be able to access available data anywhere and anytime [17]. A new ecosystem is created where sharing of information and large amount of data is possible and the capability of delivering services and demands to everyone is done with ease. With these technological innovation, smart systems had been implemented into different technological industries and automation systems. It has shown great benefits into diverse applications like agricultural innovations [18], [19], healthcare systems [20], [21], industry automation [22], [23] to the simplest applications like wearables [24], [25].

With the latest technological development of interconnecting nodes and devices, the smart home industry has expanded and grown rapidly. And with the generation of latest communication technologies, information transmission has been enhanced and improved. Smart home systems have been the common recipient of this development where everybody can experience the innovation inside their homes. In smart home environment, a collection of home appliances and devices connected to the network can be remotely controlled and automated through mobile terminals such as mobile devices [26].

Another wide scale technological implementation of data acquisition that is in current trends is smart city architecture [27]. Energy optimization within the larger area of implementation has been the most important consideration for energy sustainability. For efficient use of sustainable energy resources, sustainable energy power systems are considered together with the use of energy efficient node devices and sensors and software algorithms that enhance the productivity and storage capability of these systems. With the integration of data acquisition into smart grids, it enhanced the intelligent collection of real-time power production and environmental conditions of the systems thus increasing quality and performance of energy production, improving sustainability while reducing resource consumption [28].

In the current medical environment, data acquisition has been used to monitor human vital signs remotely. Smart and autonomous healthcare applications that performs non-invasive process of monitoring vital human health data is on the rising trend [29]. By enabling close patient monitoring, remote diagnosis is simplified while optimizing accuracy of the diagnosis and maximizing the benefits of treatment.

Data acquisition is also implemented on service-oriented industry where common requirement of users are collected so as to enhance the services provided by the industry. Improving customer service through enhanced intelligent decision-making system is one of the important factors that enterprises need to provide [30]. An intelligent assistant decision-making model can be designed to collect data form remote node devices in assisting industrial organizations in providing provide accurate and real-time updates and information to its clients.

3.1   Smart Energy Management Systems

With the expansion of global network infrastructure and the advanced development of telecommunication systems, the use of smart grids has exponentially increased [31]. In smart energy systems effective energy management, smart grids collect energy data using interconnected intelligent node devices, such as data concentrator units (DCUs) and smart meters. Integrating data acquisitions into these systems for monitoring purposes greatly improves system supervision, data acquisition and low maintenance cost.

It is crucial for renewable energy systems to be monitored in real-time because of the intermittent nature of these renewable energy sources and its sensitivity to changing environmental parameters [32]. Data acquisition systems integrated into renewable energy installations revolutionizes these technologies because of enhance efficiency, reliability and sustainability of these renewable energy systems [33]. For an energy system to be smart, an efficient energy management strategy is developed to maximize and manage power management in a micro grid [34]. This emerging information and communications technologies have a significant effect on the development of smart systems and industries including buildings, hospital, transportation and private and public infrastructure which improve the quality of systems and services [35].

One of the important characteristics of a smart energy system is its ability to analyze and adjust power consumption based on the current requirement of the system. A smart energy system collects data regarding its occupancy, lighting, temperature and power consumption of a certain room and makes data-driven decisions to improve energy usage [36]. With this functionality, the system is able to implement power control strategies like power gating and dynamic central power regulation based on real-time demand to achieve energy efficiency.

For a smart energy systems, enhanced security and privacy while optimizing energy consumption can be an added innovation with data acquisition systems. Combining machine learning with node devices for collecting real-time data on energy usage can help predict future power consumption patterns [37]. Leveraging machine learning algorithms, information from the remote node sensors can be extracted from energy consumption data and analyzed to predict future use energy requirement.

Using artificial neural network (ANN) that processes collected data from remote sensors through back propagation algorithm can also help predict future energy demand which leads to optimized energy consumption, cost savings and reduced environmental impact [38]. Sensor data from node devices such as motion sensors, temperature sensors and smart appliances can be collected and the analyzed information regarding occupancy patterns, weather conditions and energy consumption can be used determine the needed energy.

Energy consumed and its equivalent cost can also be the determining factor in the system power and electrical efficiency [39], [40]. Adding data acquisition to the system with remote node sensors and devices can help monitor real-time energy usage and the same time provide capability of controlling the system remotely. With remote control functionality, energy consumption can be minimized by allowing only priority loads to be connected if energy required is more that the available energy supply.

Meter tampering leading to energy theft has been one of the problems by the energy sector [41]. To overcome this complexity of monitoring every consumer, advanced data acquisition technology of combined smart meters and intelligent sensors are implemented. With this functionality, concerns regarding energy theft can be eliminated thereby enhancing energy management.

## 4   Future and Challenges of Data Acquisition Systems

Data acquisition systems have grown tremendously and are implemented on different applications over the past years. However, with such technological advancement it has also introduced significant privacy and security challenges and data transmission issues, including unauthorized access, data breaches, vulnerabilities in device connectivity, connectivity issues to name a few [42]. There has been a record of recent attacks in node devices such as smart bulb attack, frantic locker smart TV attack and Mirai DDoS attack on IoT. With the emergence of black hat attackers, they take the advantage in the vulnerabilities of these systems to execute cyber-attacks [43]. Different attack vectors had been identified based on its vulnerabilities, and with this, cloud providers offer multiple security techniques to mitigate if not eliminate the possibility of network attack.

Because of recent security issues, novel solutions are needed to make the network more secure and resistant to cyber and physical attacks. In the current times, since smart grid systems are already connected to networks, they become a target and cause concern on the adoption of these networked devices [44]. There had been solutions and technologies that had been presented in minimizing these challenges such as using advanced encryption methods, cloud computing technologies and advance sensor technologies [45].

Another identified challenge that is due to the vulnerabilities of node devices are malware attacks [46]. With the availability of these malware, it is important to protect devices and processes in order to prevent such attacks and secure these node devices to the threats. The need for improved malware detection either traditional or learning-based detection algorithms is needed. Using lightweight encryption technology for security and privacy of network node devices can be implemented [47]. Because latest sensor technology still cannot provide public key encryption, a lightweight cryptographic algorithm can be used to enhance the privacy of data being exchanged by these node devices.

A new set of solutions has also emerged in helping mitigate the risks on network vulnerabilities and threats for networked systems using artificial intelligence, machine learning techniques and blockchain technologies [48], [49]. The convergence of AI and blockchain technologies have revolutionized the network industry and its future sustainability to emerging security issues and challenges. Likewise, access to these node devices and collected data is secured thus preventing unauthorized control and protecting users' privacy.

Connecting energy systems to the network not only causes concern on security issues but poses several network challenges also such as bandwidth management, interfacing interoperability, connectivity, packet loss and data processing [50]. These concerns can cause problems in analyzing real-time information for proper action by the management. Since accurate and precise data is important, these issues need to be considered in the implementation.

Since the main purpose of data acquisition is to perform real-time monitoring, the requirement of always available network should be perfected [51]. A hybrid data transmission network that is to be employed as the main transmission medium for the system should be ensured so as to have no possible downtime in case of network unavailability. It can be a combination of available communication technologies such as radio [52], mobile communication [53], data transmission [54] or short-range technology [55].

## 5    Conclusion

Multiple technological improvements with regards to data acquisition implementation have emerged and grown exponentially in the last years. Integrating these node sensors to these systems have shown great improvement on the operational performance and enhance reliability of the system. The future of this technology is on the positive track with recent development of new and novel technological breakthroughs either on hardware, software or strategical techniques on the implementation of these discoveries.

The broad coverage of data acquisition implementation is limitless with increasing applications and systems trying to fit into remote control and management functionality on any existing and emerging technological system. Complexity on the design and implementation is guaranteed but the rewards to the improvement on remote management and control of the system overshadow this.

As data acquisition systems become an integral part in the renewable energy platform, predictive maintenance in monitored systems enhances the management strategies as well as improves the operation, safety and reliability of energy generation systems. With the introduction of Internet of Everything, the newer phase of evolution for Internet of Things, a model for interconnecting machine-to-machine communications, data acquisition technologies have greatly advanced.

## 6    Recommendations

Since data acquisition implementation has been into different applications and systems, it has also garnered emerging technological challenges specifically in data transmission security and privacy. The need for further study on these issues must be given consideration to enhance the integrity and accuracy of data being exchanged by these systems. Maximizing the use of cloud computing will greatly improve big data processing capability and enhance integrity and improve accuracy of interpreted information.

With the current design of available network infrastructure, the need for further consideration on different technological enhancement on network implementation is a must so as to fully support the always availability requirement of any data acquisition systems. Latest telecommunication standards and protocols can be explored further to achieve multiple options on possible ways to transmit and receive data without interruption.

Considering artificial intelligence and advanced algorithms for detecting and identifying data manipulation and compromised data integrity would greatly improve the decision making capability and management strategies of such systems. With current technologies for security threats elimination and malicious attacks prevention, operational disruptions and incorrect decisions are prevented.

# 7 Acknowledgements